\documentclass[11pt]{article}
\usepackage{moriond,epsfig}
\usepackage{amsmath,amssymb}
\usepackage{multirow}

\def\MyPhoto{\psfig{figure=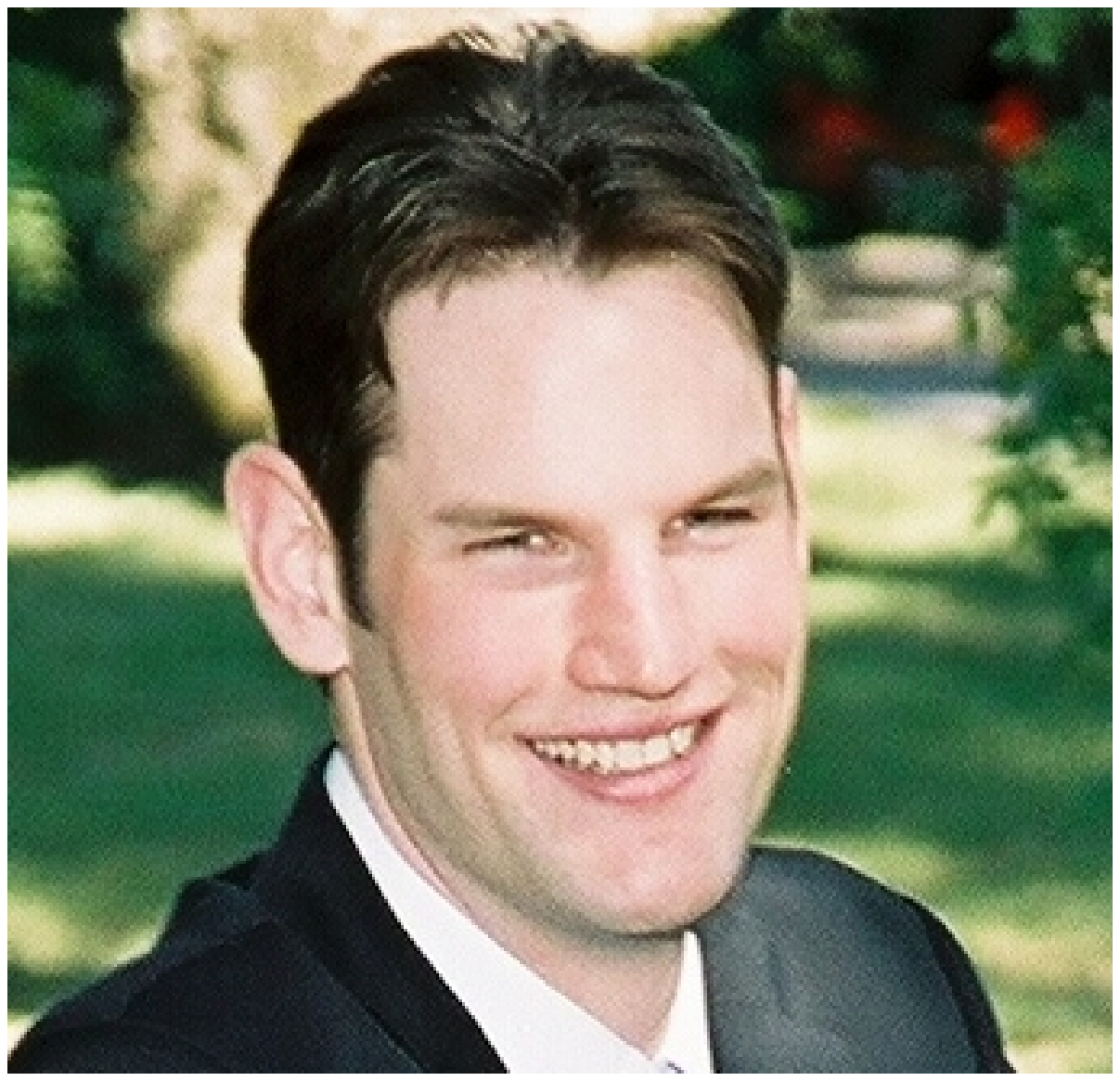,width=35mm}}

\def\be{\begin{equation}}
\def\ee{\end{equation}}
\def\bea{\begin{eqnarray}}
\def\eea{\end{eqnarray}}

\def\evqq{\ensuremath{e\nu q\bar q}}
\def\mvqq{\ensuremath{\mu\nu q\bar q}}
\def\wjets{$W+$jets}
\def\zjets{$Z+$jets}
\def\ttbar{$t\bar{t}$}
\def\met{\ensuremath{E\kern-0.57em/_{T}}}
\def\mida{$+2/-3$}
\def\midb{$+0/-5$}

\begin{document}
  \vspace*{4cm}
  \title{FIRST EVIDENCE OF $WW/WZ\to \ell\nu q\bar q$ AT THE TEVATRON}
  
  \author{ JOSEPH HALEY \\ (For the D0 Collaboration)}
  
  \address{Princeton University Department of Physics\\
    Princeton, NJ, USA}
  
  \maketitle
  
  \abstracts{ We present the first evidence from a hadron collider of
    $WW+WZ$ production with semi-leptonic decays. The data were
    recorded by the D0 detector at the Fermilab Tevatron and
    correspond to 1.07~fb$^{-1}$ of integrated luminosity obtained in
    proton-antiproton collisions at $\sqrt{s}=$1.96~TeV.  The cross
    section observed for $WW+WZ$ production is $20.2\pm 4.5$~pb with a
    significance of 4.4 standard deviations.  }

  \section{Introduction}
  
  There are many reasons for studying $WW/WZ\to \ell\nu q\bar q$ at
  the Tevatron.  From the electroweak prospective, diboson production
  provides a probe of self-interactions of vector bosons.  Deviations
  from the Standard Model (SM) of these trilinear gauge boson coupling
  would affect the cross sections and event kinematics of diboson
  production~\cite{bib:ac}.  The cross sections for diboson production
  at the Tevatron had previously only been measured for the fully
  leptonic final states~\cite{bib:dz,bib:cdf}, so this analysis
  provides a compliment to the previous measurements.

  Reconstruction of $WW$ and $WZ$ events in semi-leptonic final states
  represents a challenge in separating signal from the dominant
  background of a $W$ boson produced in association with jets.  This
  is a challenge shared by many Higgs boson searches, {\it e.g.}
  $WH\to\ell\nu b\bar b$, making this measurement a benchmark for
  these similar Higgs boson searches.  Furthermore, this analysis
  provides a proving ground for the multivariate event-classification
  schemes and the accompanying statistical
  techniques~\cite{bib:poisson} that are used for the Tevatron Higgs
  boson searches in the entire mass range allowed by the SM.

 \section{Event Selection}

\begin{figure}[htbp]
    \vspace{2mm}
    \centering
    \begin{tabular}{ccc}
      (a)&(b)&(c)\\
      \includegraphics[width=1.95in]{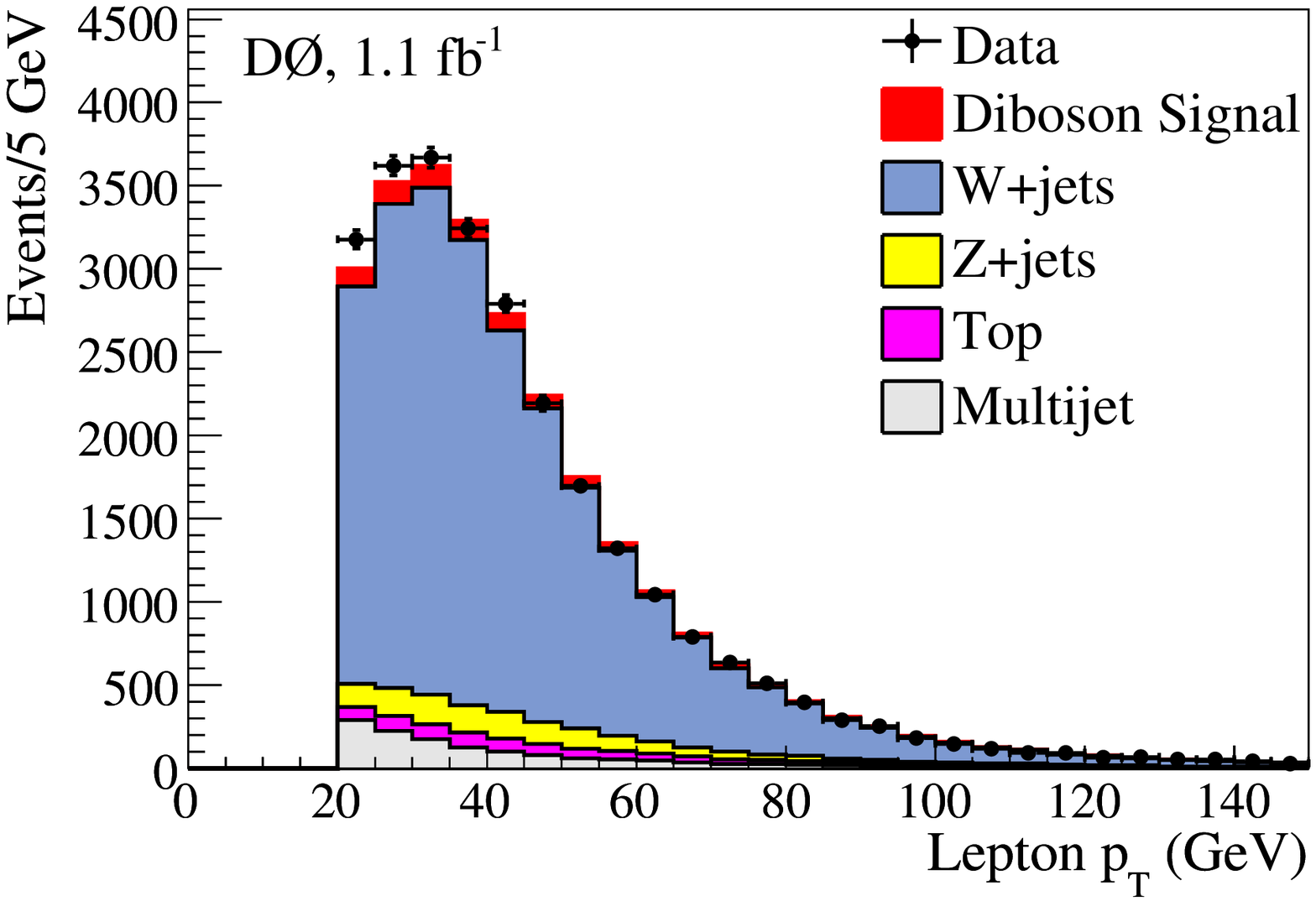}&
      \includegraphics[width=1.95in]{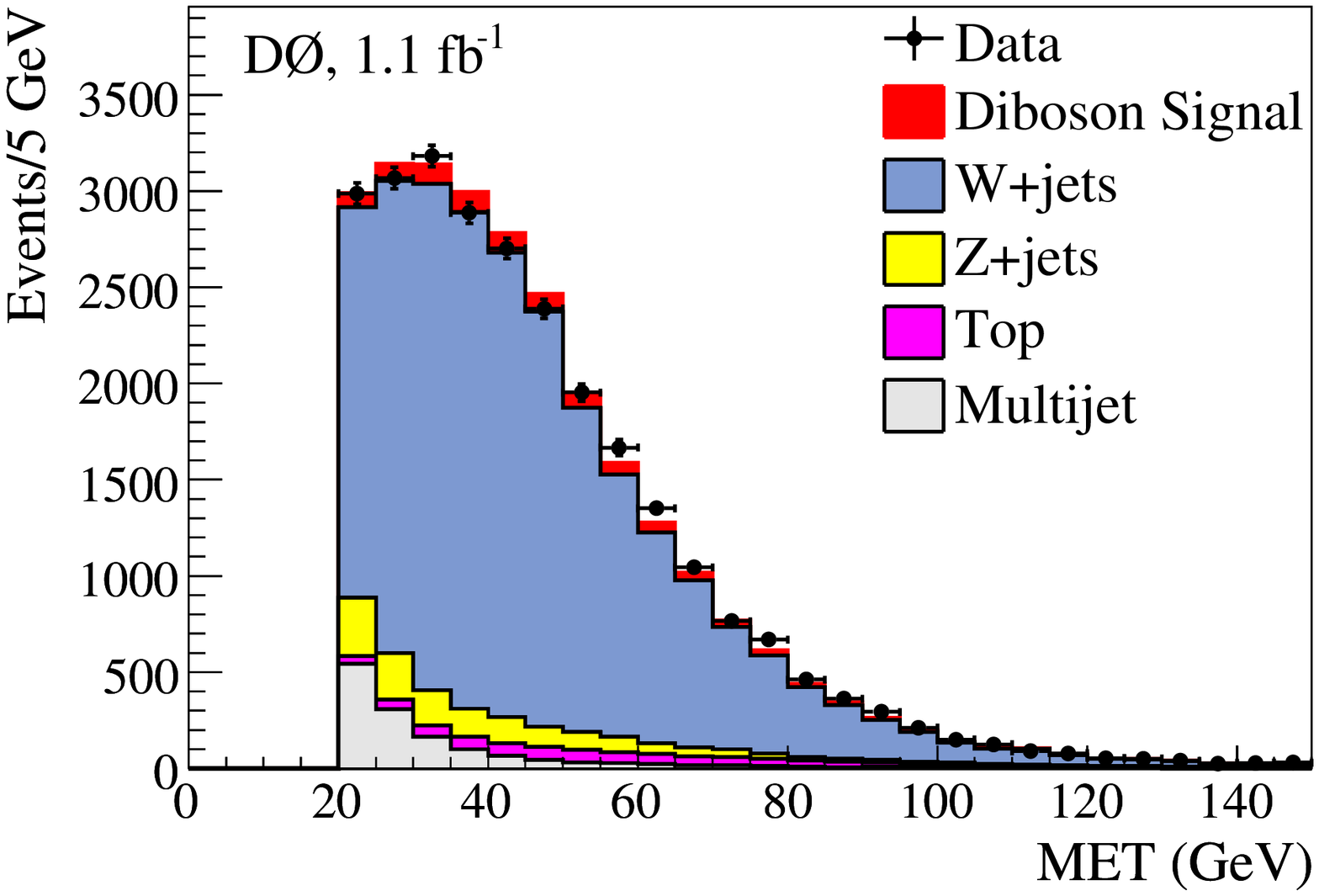}&
      \includegraphics[width=1.95in]{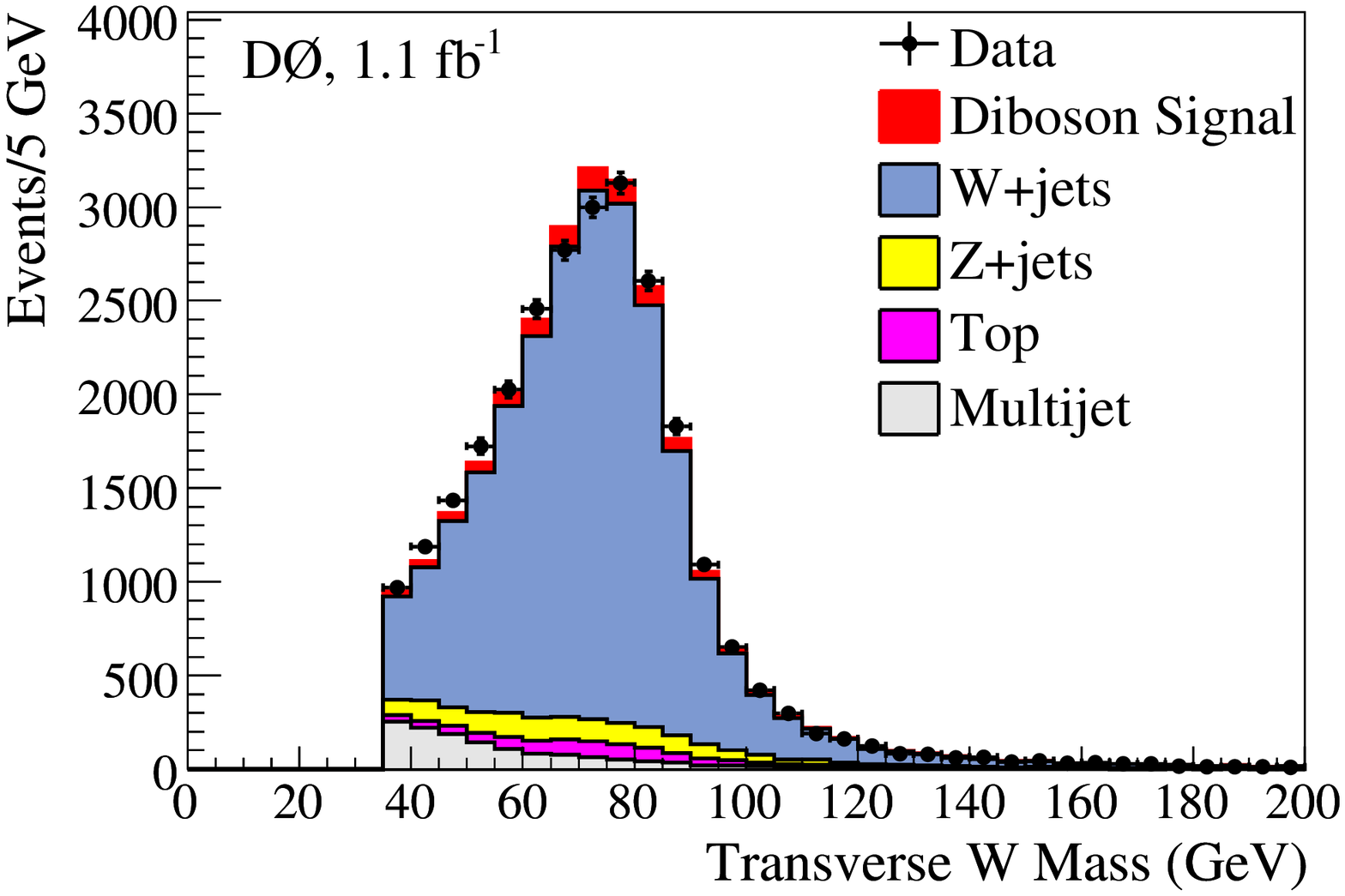}\\
      (d)&(e)&(f)\\
      \includegraphics[width=1.95in]{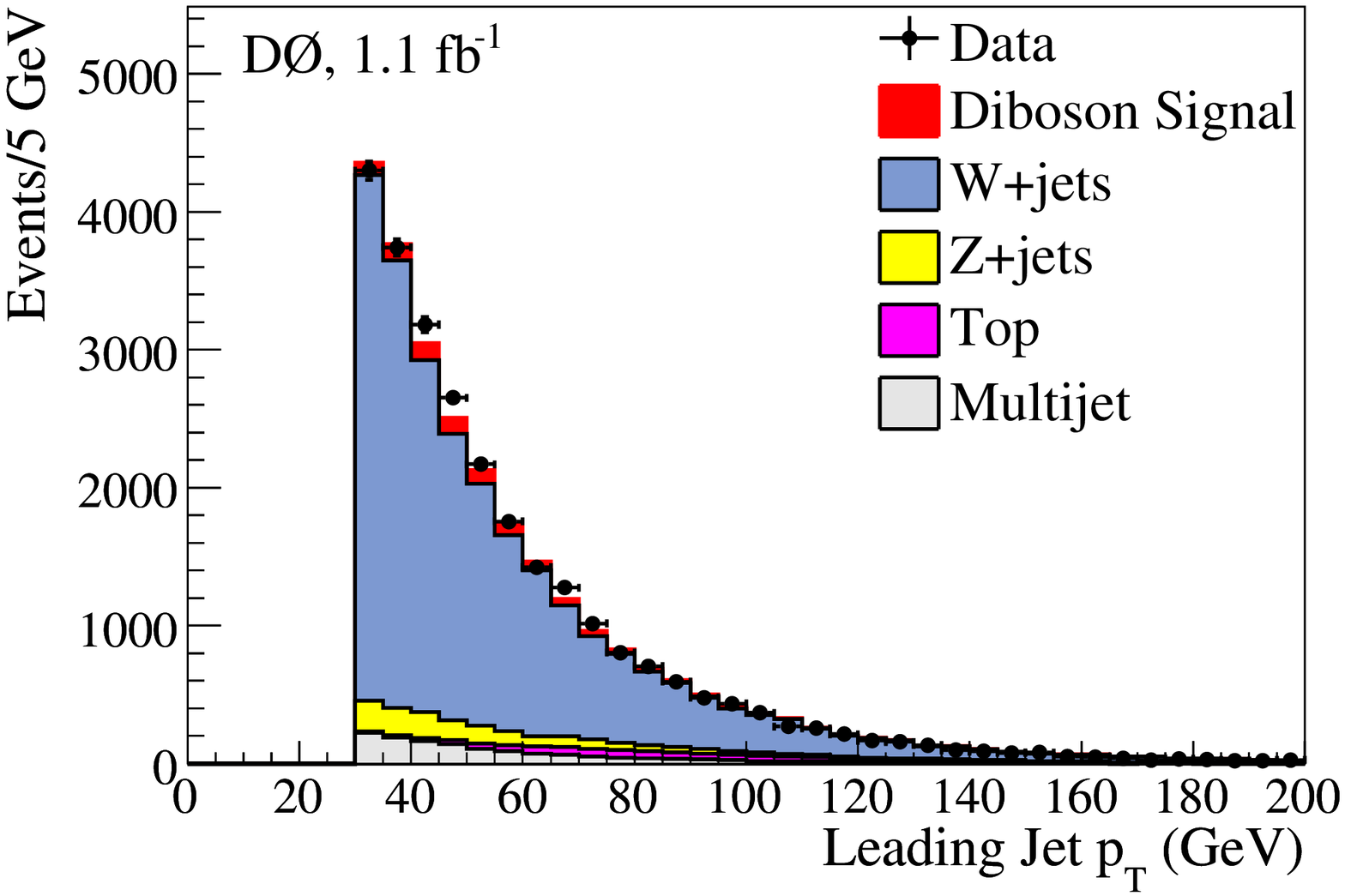}&
      \includegraphics[width=1.95in]{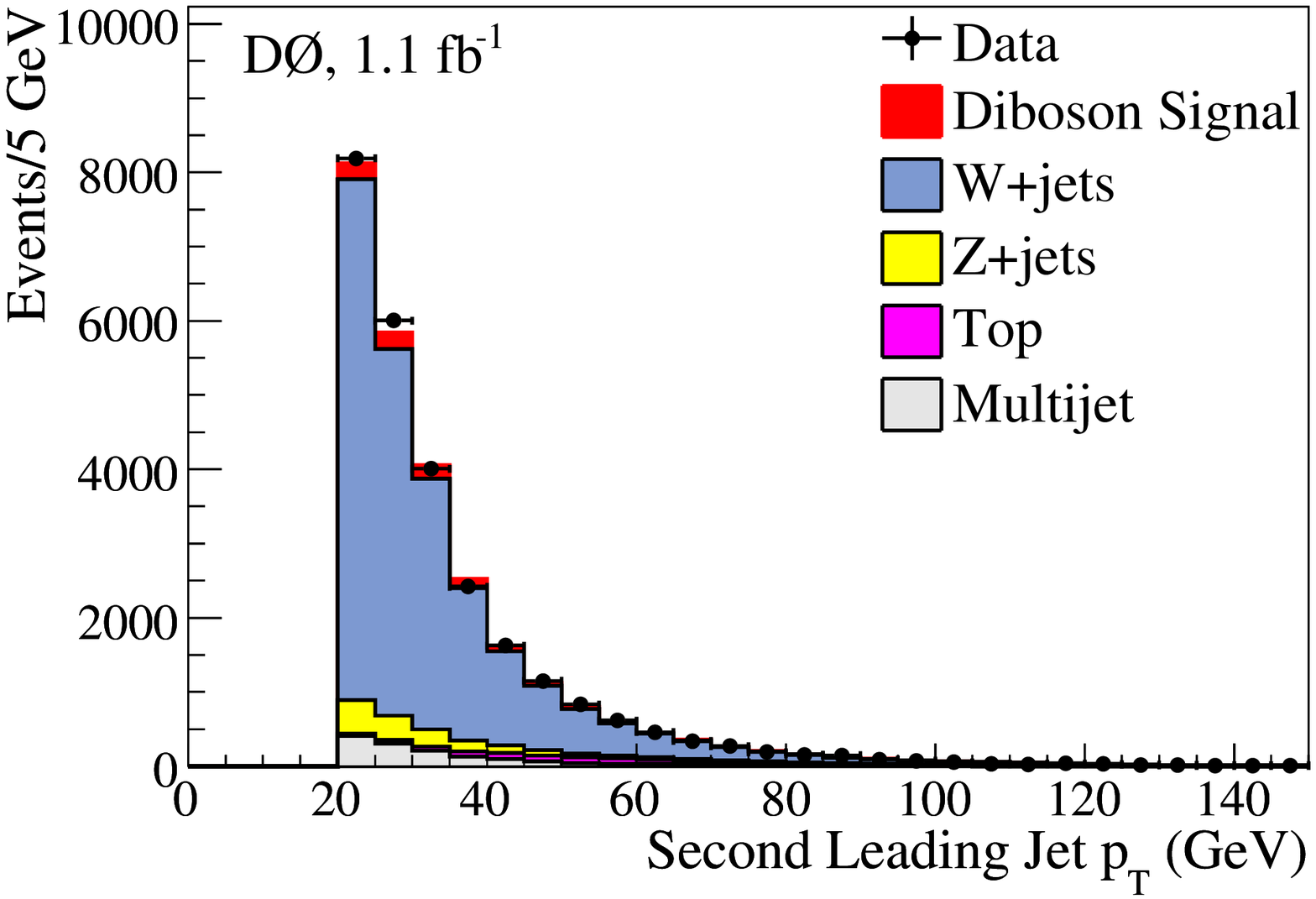}&
      \includegraphics[width=1.95in]{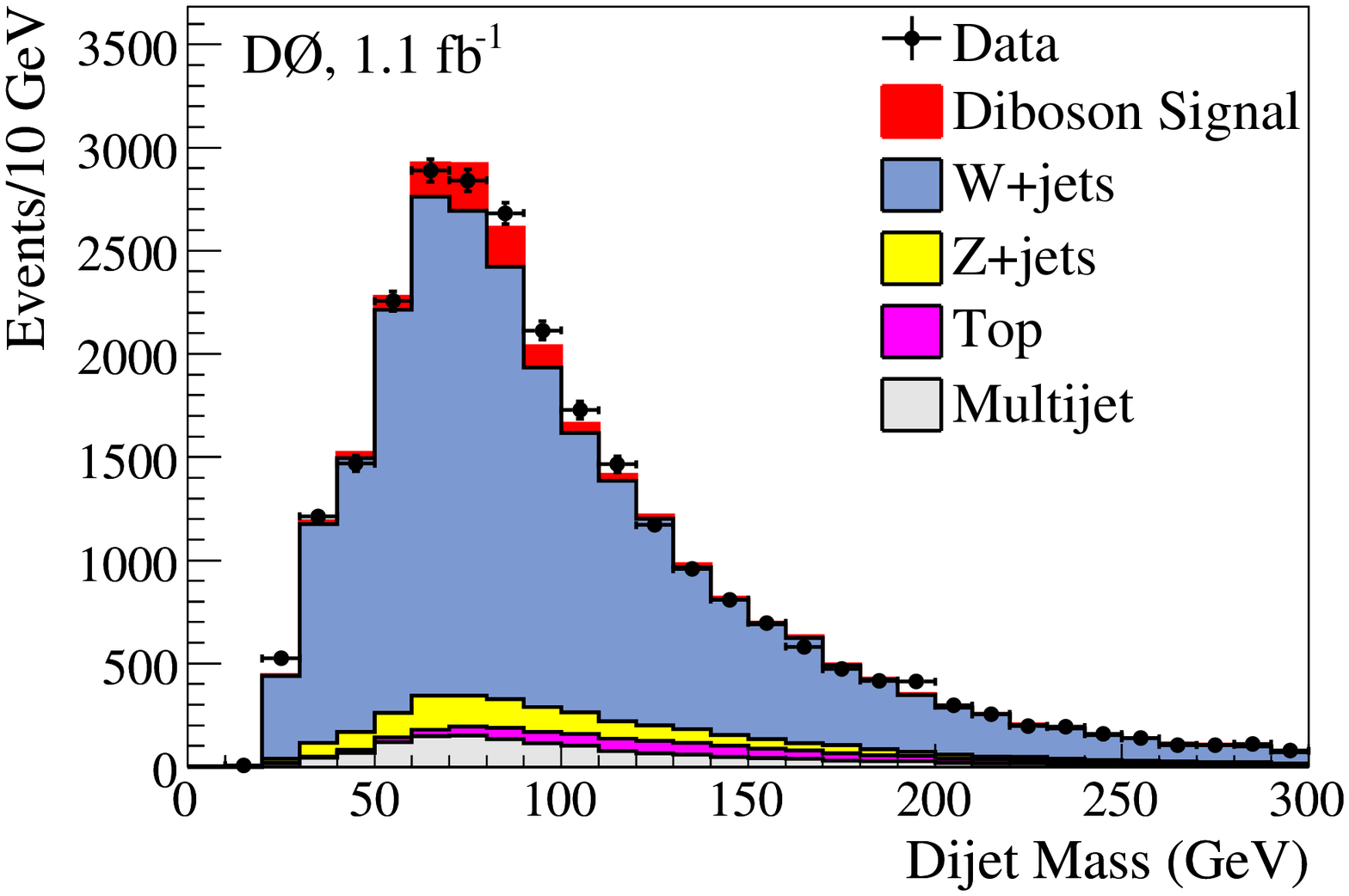}\\
    \end{tabular}
    \caption{Kinematic distributions after all selection requirement:
      (a) $p_T$ of lepton; (b) \met; (c) transverse W mass; (d) $p_T$
      of leading jet; (e) $p_T$ of second-leading jet; (f) dijet
      mass. }
    \label{fig:kin}
  \end{figure}

  To select candidate events for $p\bar{p}\to WW/WZ\to \ell\nu q\bar
  q$, we required a single reconstructed lepton (electron or
  muon)~\cite{bib:leptons} with transverse momentum $p_T>20$~GeV and
  $|\eta|<1.1$ (for electrons) or $|\eta|<2$ (for muons), an imbalance
  in transverse momentum \met$>20$~GeV, and at least two
  jets~\cite{bib:JetCone} with $p_T>20$~GeV and $|\eta|<2.5$.  The
  leading jet ({\it i.e.} with the highest $p_T$) was also required to
  have $p_T>30$~GeV.  To reduce background from processes that do not
  contain $W\to\ell\nu$, we required a transverse $W$ mass of $M_T^W>$
  35~GeV, where $M_T \equiv
  \sqrt{(E_T)^2-(\vec{p}_T)^2}$~\cite{bib:smithUA1}.  The electron or
  muon trajectories were required to be isolated from other objects in
  the calorimeter, and had to match a track reconstructed in the
  central tracking system that originated from the primary vertex.
  Also, the muon had to be reconstructed as an isolated track in the
  central tracking system.  The resulting kinematic distributions are
  shown in Fig.~\ref{fig:kin}.

  \section{Data Sample}
  
  The data were collected with the D0 detector~\cite{bib:detector} at
  the Fermilab Tevatron Collider at a center-of-mass energy of
  $\sqrt{s}=$ 1.96~TeV.  The events studied in this analysis
  correspond to 1.07~fb$^{-1}$ of integrated luminosity collected
  during Run IIa (2002-2006).  To be considered for analysis, events
  in the $e\nu q\bar{q}$ channel were required to pass at least one
  single electron or electron+jet(s) trigger.  The resulting trigger
  efficiency was $98^{+2}_{-3}$\%.  A suite of triggers was used for
  the $\mu\nu q\bar{q}$ channel resulting in a trigger efficiency of
  nearly 100\%.

  \section{Signal and Background Estimations}

  Monte Carlo generators were used to simulate the signal and
  background samples that contained a charged lepton in the final
  state.  Signal events were generated with {\sc
    pythia}~\cite{bib:PYTHIA} using CTEQ6L parton distribution
  functions (PDF).  {\sc Alpgen}~\cite{bib:ALPGEN} with CTEQ6L1 PDFs
  was used to generate \wjets, \zjets, and \ttbar\ events and {\sc
    comphep}~\cite{bib:CompHEP} with CTEQ6L1 PDFs was used to simulate
  single-top events.  All {\sc alpgen} and {\sc comphep} events used
  {\sc pythia} for parton showering and hadronization.  After
  generation, the events underwent a {\sc
    geant}-based~\cite{bib:GEANT} detector simulation before being
  reconstructed with the same programs as the data.
  
  With the exception of \wjets, all background MC samples were
  normalized to next-to-leading-order (NLO) or
  next-to-next-to-leading-order SM predictions.  The
  \wjets\ normalization was determined simultaneously with the signal
  cross section by a fit to data, as discussed later.

  The probability for a multijet event to mimic a lepton and pass all
  selection cuts was quite small; however, because the cross section
  for multijet production is so large, the background from multijet
  events had to be accounted for.  For the \mvqq\ channel, the multijet
  background was modeled with ``anti-isolated'' data corresponding to
  events that failed the muon isolation requirements, but passed all
  other selections.  The kinematic distributions of the anti-isolated
  data were corrected for contributions from processes already modeled
  via MC. The normalization of the multijet background in the muon
  channel was determined from a fit to the transverse $W$ mass
  distribution of the $\mu+\nu$ system.

  For the \evqq\ channel, the multijet background was estimated using
  a ``loose-but-not-tight'' (LNT) data sample obtained by selecting
  events that passed a loosened electron-quality requirement, but did
  not pass the electron-quality requirement of the final
  selection~\cite{bib:leptons}. To estimate the correct rate for
  multijet events, a weight was applied to each LNT event based on the
  probability for a jet to mimic an electron.  Also, the contribution
  from events that were already modeling via MC was subtracted.
  
  \begin{table}[tbp] 
    \caption{Measured number of events for signal and each background
      after the combined fit of the RF distribution (with total
      uncertainties determined from the fit) and the observed number
      of selected events.}
    \label{tab:yields}
    
    \centering
    \begin{tabular}{| l @{ \extracolsep{\fill} } r @{$\ \pm\ $\extracolsep{0cm}} l @{ \extracolsep{\fill} } r @{$\ \pm\ $\extracolsep{0cm}} l |}
      \hline
      & \multicolumn{2}{c}{$e\nu q\bar{q}$ channel} & \multicolumn{2}{c |}{$\mu\nu q\bar{q}$ channel} \\
      \hline	   
      Diboson signal        &   436  &  36  &   527  &  43 \\
      $W$+jets              & 10100  & 500  & 11910  & 590 \\
      $Z$+jets              &   387  &  61  &  1180  & 180 \\
      \ttbar\ + single top  &   436  &  57  &   426  &  54 \\
      Multijet              &  1100  & 200  &   328  &  83 \\
      \hline
      Total predicted       & 12460  & 550  &  14370 & 620 \\
      Data                  & \multicolumn{2}{c}{12473} & \multicolumn{2}{c |}{14392} \\
      \hline
    \end{tabular}
  \end{table}

  \section{MC Corrections and Systematic Uncertainties}

  As one can see from Table~\ref{tab:yields}, contributions to the
  selected events was dominated by the background from \wjets.
  Therefore, accurate modeling of the \wjets\ background was of
  particular importance.  We performed detailed studies of the {\sc
    alpgen} \wjets\ MC sample and associated sources of uncertainty.
  Comparison with other generators and data showed discrepancies
  between the modeling of jet $\eta$ and $\Delta R$ between
  jets~\cite{bib:ALPGENcomp}.  Therefore, the data were used to
  correct these quantities in the {\sc alpgen} \wjets\ and
  \zjets\ samples.  The effect of the diboson signal on the derived
  corrections was small, but nonetheless taken into account via a
  systematic uncertainty assigned to the procedure.  The {\sc alpgen}
  \wjets\ sample was also assigned systematic uncertainties for
  variations of the renormalization (and factorization) scale and
  jet-parton matching parameters~\cite{bib:MLM}.  PDF uncertainties
  were evaluated for all of the MC samples, as were uncertainties from
  object reconstruction and identification.  A full list of the
  systematic uncertainties and the magnitude of each is given in
  Table~\ref{tab:systEMMU}.  We considered systematic uncertainties
  that affected both normalization and the shapes of kinematic
  distributions.

  \begin{table}[htbp] 
    
    \caption{The \% systematic uncertainties for Monte Carlo
      simulations and multijet estimates. Uncertainties are identical
      for both lepton channels except where indicated otherwise. The
      nature of the uncertainty, i.e., whether it had a differential
      dependence (D) or just normalization (N), is also provided. The
      values for uncertainties with a differential dependence
      correspond to the RMS amplitudes in the RF output distribution.
      Also provided is the contribution of each source to the total
      systematic uncertainty of 3.6~pb on the measured cross section.
    }
      
    \label{tab:systEMMU}

    \footnotesize\selectfont
    \centering
    \begin{tabular}{| l @{\extracolsep{\fill}} 
        r @{\ \ \extracolsep{\fill}} l @{\extracolsep{\fill}\ \ } 
        r @{\ \ \extracolsep{\fill}} l @{\extracolsep{\fill}\ \ } 
        r @{\ \ \extracolsep{\fill}} l @{\extracolsep{\fill}\ \ } 
        r @{\ \ \extracolsep{\fill}} l @{\extracolsep{\fill}\ \ } 
        r @{\ \ \extracolsep{\fill}} l @{\extracolsep{\fill}\ \ } 
        c |}
      \hline
      \multicolumn{1}{| c}{Source of systematic}
      & \multicolumn{2}{c}{\multirow{2}{*}{Diboson}}
      & \multicolumn{2}{c}{\multirow{2}{*}{$W$+jets}}
      & \multicolumn{2}{c}{\multirow{2}{*}{$Z$+jets}}
      & \multicolumn{2}{c}{\multirow{2}{*}{Top}}
      & \multicolumn{2}{c}{\multirow{2}{*}{Multijet}}
      & \multicolumn{1}{c |}{\multirow{2}{*}{$\Delta\sigma\;(pb)$}}\\
      \multicolumn{1}{| c}{uncertainty}
      &&
      &&
      &&
      &&
      &&
      &\\
      
      \hline
      Trigger efficiency, $e\nu q\bar{q}$ channel          &&   \mida  &&  \mida   &&  \mida   &&  \mida   &&         & $<0.1$\\
      Trigger efficiency, $\mu\nu q\bar{q}$ channel        &&   \midb  &&  \midb   &&  \midb   &&  \midb   &&         & $<0.1$\\
      Lepton identification                                &&  $\pm$4  &&  $\pm$4  &&  $\pm$4  &&  $\pm$4  &&         & $<0.1$\\
      Jet identification                                   &&  $\pm$1  &&  $\pm$1  &&  $\pm$1  &&  $\pm<$1  &&        & 0.3\\
      Jet energy scale                                     &&  $\pm$4  &&  $\pm$9  &&  $\pm$9  &&  $\pm$4  &&         & 1.9\\
      Jet energy resolution                                &&  $\pm$3  &&  $\pm$4  &&  $\pm$4  &&  $\pm$4  &&         & $<0.1$\\
      
      Cross section                                        &&          && 
      
      $\pm$20
      
                                                                                   &&  $\pm$6  &&  $\pm$10 &&         & 1.1\\
      Multijet normalization, $e\nu q\bar{q}$ channel      &&          &&          &&          &&          && $\pm$20 & 0.9\\
      Multijet normalization, $\mu\nu q\bar{q}$ channel    &&          &&          &&          &&          && $\pm$30 & 0.5\\
      Multijet shape, $e\nu q\bar{q}$ channel              &&          &&          &&          &&          && $\pm$6  & $<0.1$\\
      Multijet shape, $\mu\nu q\bar{q}$ channel            &&          &&          &&          &&          && $\pm$10 & $<0.1$\\
      Diboson signal NLO/LO shape                          &&  $\pm$10 &&          &&          &&          &&         & $<0.1$\\
      Parton distribution function                         &&  $\pm$1  &&  $\pm$1  &&  $\pm$1  &&  $\pm$1  &&         & 0.2\\ 
      {\sc alpgen} $\eta$ and $\Delta R$ corrections       &&          &&  $\pm$1  &&  $\pm$1  &&          &&         & $<0.1$\\
      Renormalization and factorization scale              &&          &&  $\pm$3  &&  $\pm$3  &&          &&         & 0.9\\ 
      {\sc alpgen} parton-jet matching parameters          &&          &&  $\pm$4  &&  $\pm$4  &&          &&         & 2.4\\ 
      \hline
    \end{tabular}
  \end{table}

  \section{Multivariate Classification}

  Improved separation between the signal and the backgrounds was
  achieved using a multivariate classification technique to combine
  information from several kinematic variables. The technique used was
  a random forest (RF) classifier~\cite{bib:SPR1,bib:SPR2} from the
  {\sc StatPatternRecognition}~\cite{bib:SPR2} software package.  The
  RF algorithm creates many decision tree classifiers, which are
  basically a series of optimized binary splits to separate signal
  from background.  The RF is then formed by taking the average of all
  of the decision trees.  The key to the RF is that each decision tree
  uses only a subset of the input variables (selected randomly for
  each tree) and is trained on a bootstrap replica~\cite{bib:SPR1} of
  the full training set.  This results in each of the trees
  generalizing differently to unseen data because each tree was
  trained with differently.  The net effect of then averaging all the
  trees is an accurate and stable classifier.

  The inputs to the RF were thirteen well-modeled kinematic variables
  that demonstrated a difference in probability density between signal
  and at least one of the backgrounds.  A RF for each channel was
  trained using one half of each MC sample.  The other halves, along
  with the multijet background samples, were used to evaluate the RF
  output distributions for comparison to the data.  These RF output
  distributions were then used to measure the excess of events in the
  data consistent with the kinematics of $WW$ and $WZ$ production
  (over that expected from multijet and other SM processes).

  \section{Cross Section Measurement}

  The cross section for $WW+WZ$ production was determined from a fit
  of signal and background RF templates to the data by minimizing a
  Poisson $\chi^2$ function within variations of the systematic
  uncertainties~\cite{bib:poisson}.  The systematic uncertainties were
  treated as Gaussian-distributed uncertainties on the expected
  numbers of signal and background events in each bin of the RF
  distribution.  Each individual uncertainty was treated as 100\%
  correlated between channels, samples, and from bin to bin.
  Different sources of uncertainty were assumed to be independent.
  
  The normalizations of the RF templates for the signal and the
  \wjets\ background were unconstrained in the fit; allowing the fit
  to simultaneously measured the signal cross section and determine
  the normalization of the dominant background.  This approach
  eliminated the need to use the \wjets\ cross section predicted by
  {\sc alpgen} and provided an unbiased uncertainty for the
  normalization of the dominant background.  As a check of the
  procedure, the fit yielded an effective k-factor of $1.53 \pm 0.13$
  that needed to be applied to the {\sc alpgen} cross section to best
  match the data, which is close to what one would expected from the
  ratio of NLO to LO predictions for the \wjets\ cross section.

\begin{figure}[htbp]
    \vspace{2mm}
    \centering
    \begin{tabular}{ccc}
      (a)&(b)&(c)\\
      \includegraphics[width=1.95in]{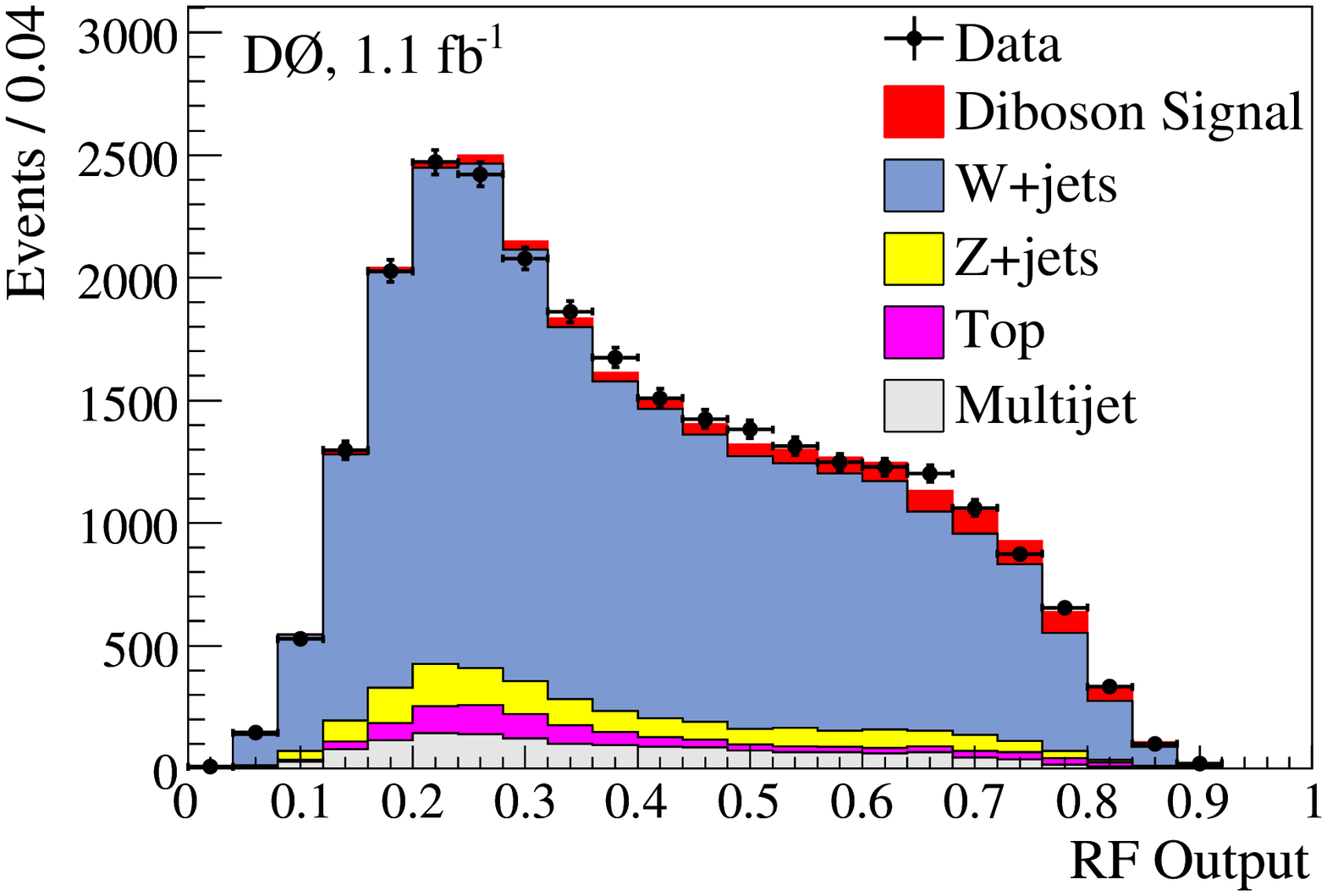} &
      \includegraphics[width=1.95in]{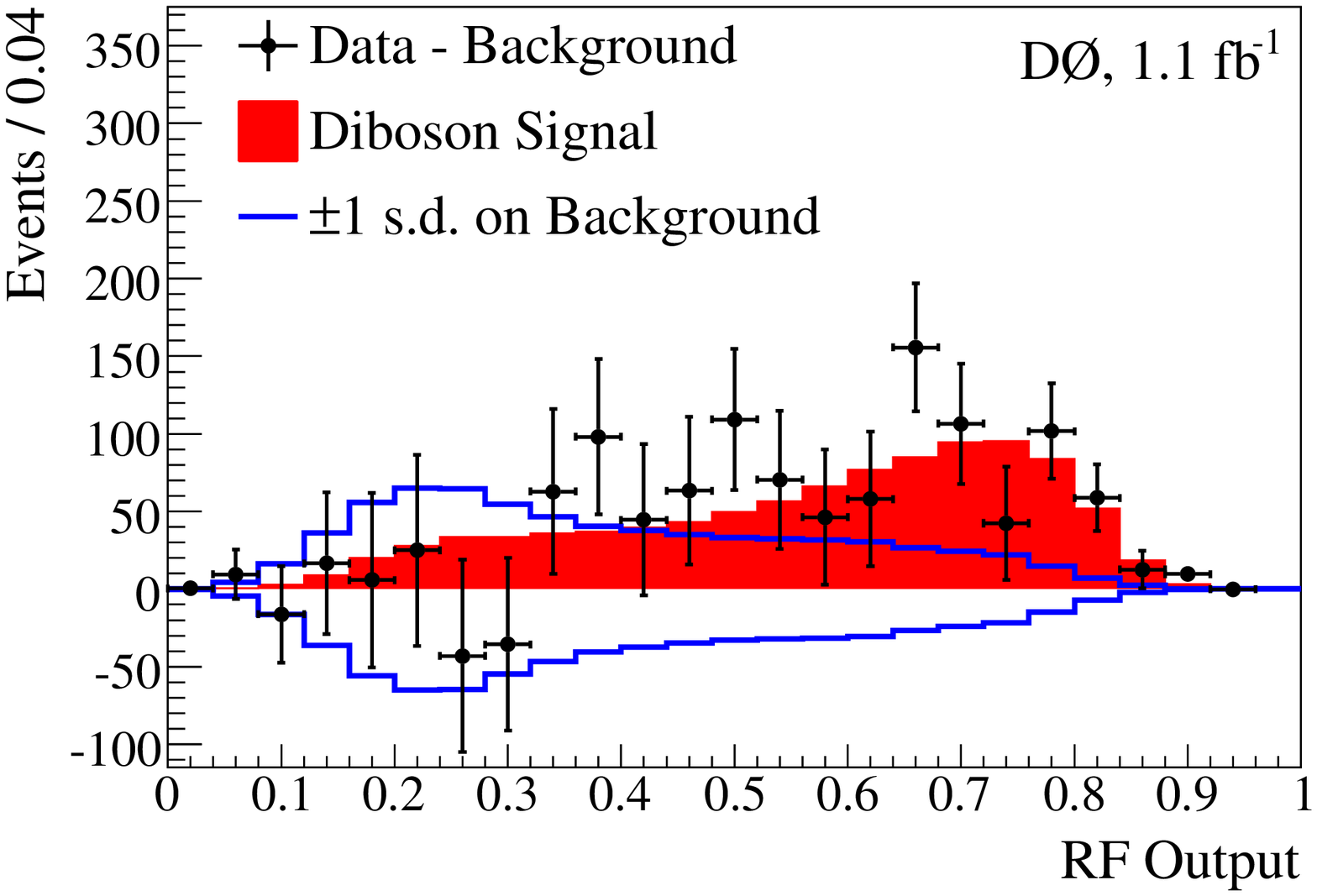} &
      \includegraphics[width=1.95in]{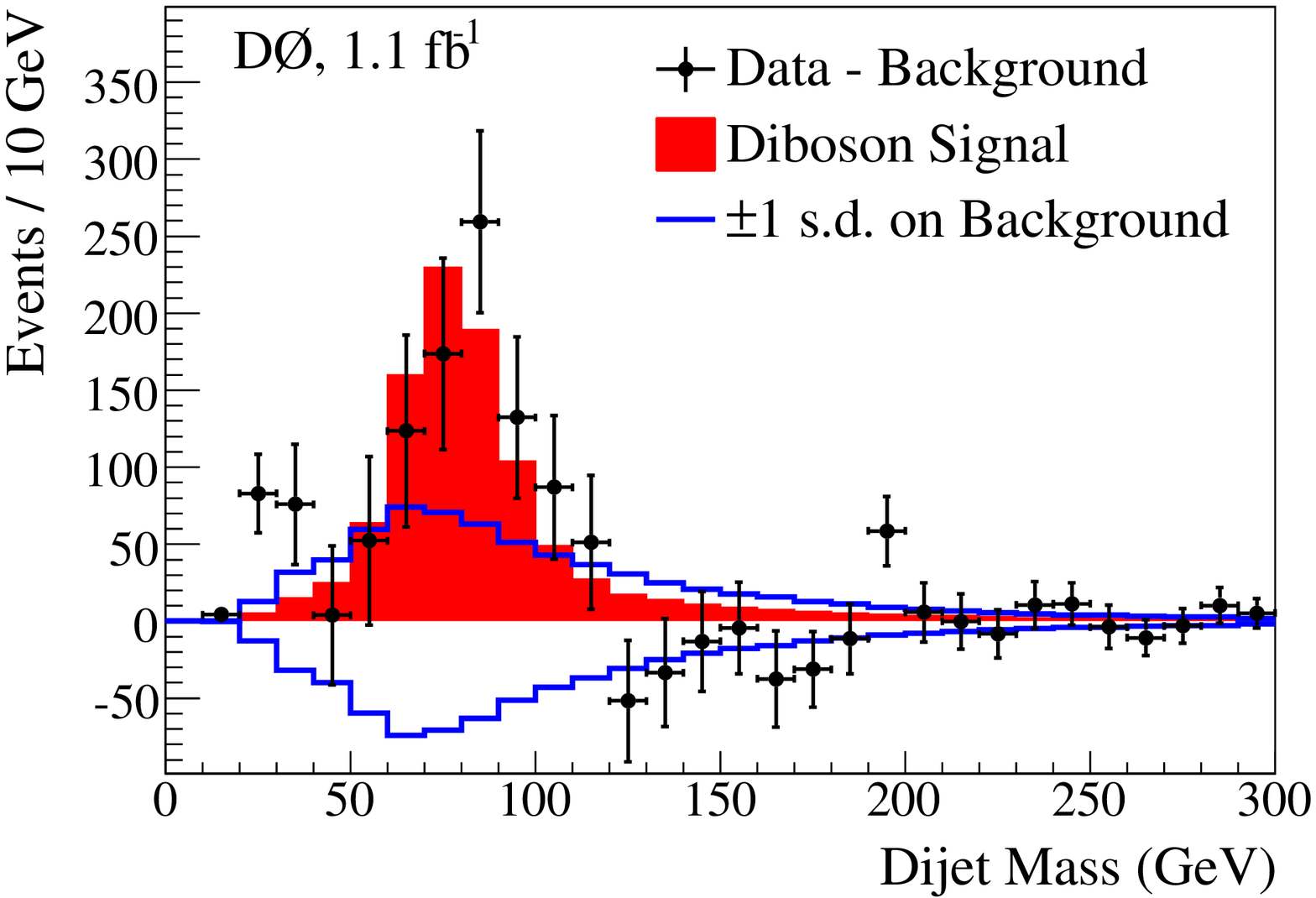} \\
    \end{tabular}
    \caption{The distributions after cross section fit of the RF
      distribution: (a) RF output; (b) RF output with background
      subtracted; (c) dijet mass with background subtracted.}
    \label{fig:fit}
  \end{figure}

  Table~\ref{tab:xsecFit} contains the results of the fit in the
  \evqq, \mvqq, and the combined channels.  The combined distribution
  of the RF output after the combined fit and the same plot with the
  background subtracted are shown in Fig.~\ref{fig:fit}.  Also in
  Fig.~\ref{fig:fit} is the background-subtracted plot for the dijet
  mass distributions showing the resonant dijet signal peak observed
  in data.  The common behavior of each fit indicates a $WW+WZ$ cross
  section consistent with, though somewhat larger than, the expected
  SM value of $\sigma(WW+WZ)=16.1$~pb~\cite{bib:Campbell}.  The
  combined lepton channel cross section fit yielded a total value of
  20.2~$\pm$~2.5(stat)~$\pm$~3.6(sys)~$\pm$~1.2(lum)~pb, which is
  slightly less that one standard deviation from expectation.

  Table~\ref{tab:xsecFit} also provides the result from preforming
  the measurement using only the dijet mass distribution.  As
  expected, the measurement from the dijet mass distribution was less
  precise than from the RF because the RF was better at discriminating
  signal from background.

  \begin{table}[htb]

    \caption{ The signal cross section determined from a simultaneous
      fit to the data of the $WW+WZ$ cross section and the
      normalization factor for \wjets. }  \centering
    \label{tab:xsecFit}
      \begin{tabular}{| lc |}
        \hline
	Channel & Fitted signal $\sigma$ (pb)\\
	\hline
	 $e\nu q\bar{q}$ RF Output   & 18.0$\pm$3.7(stat)$\pm$5.2(sys)$\pm$1.1(lum)\\
         $\mu\nu q\bar{q}$ RF Output & 22.8$\pm$3.3(stat)$\pm$4.9(sys)$\pm$1.4(lum)\\
	 Combined RF Output          & 20.2$\pm$2.5(stat)$\pm$3.6(sys)$\pm$1.2(lum)\\
	 \hline
	 Combined Dijet Mass         & 18.5$\pm$2.8(stat)$\pm$4.9(sys)$\pm$1.1(lum) \\
         \hline
      \end{tabular}
  \end{table}

  \section{Significance}

  Arguably just as important as the cross sections measurement is the
  significance of the measurement.  The expected and observed
  significances were obtained via fits of the signal plus background
  hypothesis to MC events drawn from the background-only
  hypothesis~\cite{bib:singleTop}.  The pseudo-data samples were
  generated from random Poisson trials seeded by the predicted number
  of background events smeared within the systematic uncertainties.  A
  measurement of the signal cross section was performed on each of the
  background-only pseudo-data distributions just as for the data.  The
  \emph{expected} significance corresponds to the fraction of outcomes
  that yielded a cross section at least as large as the SM prediction
  for $WW+WZ$ production.  The \emph{observed} significance was
  determined by the fraction of outcomes above the measured cross
  section.

  Table~\ref{tab:significance} gives the probability (p-value) and
  Gaussian significance (number of standard deviations for the
  corresponding Gaussian confidence level) for expected and observed
  outcomes corresponding to the measurements in
  Table~\ref{tab:xsecFit}.  Again one can see the merit of the
  multivariate classifier.  While the observed significance using the
  dijet mass was found to be 3.3 standard deviation, the RF had an
  observed significance of 4.4 standard deviations.

  \begin{table}[htb]

    \caption{ Expected and observed p-values obtained by comparing the
      measurement with background-only pseudo-experiments and the
      corresponding significance in number of standard deviations
      (s.d.) for a one-sided Gaussian integral.}  \centering
    \label{tab:significance}
      \begin{tabular}{| lcc |}
        \hline
	Channel &Expected p-value (significance)&Observed p-value (significance) \\ 
	\hline
	 $e\nu q\bar{q}$ RF Output    & $6.8\times10^{-3}$ (2.5 s.d.) & $3.2\times10^{-3}$ (2.7 s.d.) \\
         $\mu\nu q\bar{q}$ RF Output  & $1.8\times10^{-3}$ (2.9 s.d.) & $5.2\times10^{-5}$ (3.9 s.d.) \\
	 Combined RF Output           & $1.5\times10^{-4}$ (3.6 s.d.) & $5.4\times10^{-6}$ (4.4 s.d.) \\
	 \hline
	 Combined Dijet Mass          & $1.7\times10^{-3}$ (2.9 s.d.) & $4.4\times10^{-4}$ (3.3 s.d.) \\
         \hline
      \end{tabular}
  \end{table}

  \section{Conclusions}

  Using semi-leptonic decay channels, we measured $\sigma(WW+WZ) =
  20.2 \pm 4.5$~pb in proton-antiproton collisions
  $\sqrt{s}=1.96$~TeV.  This is consistent with the SM prediction of
  $\sigma(WW+WZ) = 16.1\pm 0.9$~pb as well as with previous
  measurements of $WW$ and $WZ$ in the fully leptonic final
  states~\cite{bib:dz,bib:cdf}.  The significance of the measurement
  is 4.4 standard deviations about the background, indicating the
  first direct evidence for $WW+WZ$ production with semi-leptonic
  decays at a hadron collider.  Finally, this analysis demonstrates
  the ability to measure a small signal in a large background for a
  final state of direct relevance to searches for a low mass Higgs
  boson and provides a validation of the analytical methods used in
  searches for Higgs bosons at the Tevatron~\cite{bib:tevcombo}.

  \section*{Acknowledgments}
%
We thank the staffs at Fermilab and collaborating institutions, 
and acknowledge support from the 
DOE and NSF (USA);
CEA and CNRS/IN2P3 (France);
FASI, Rosatom and RFBR (Russia);
CNPq, FAPERJ, FAPESP and FUNDUNESP (Brazil);
DAE and DST (India);
Colciencias (Colombia);
CONACyT (Mexico);
KRF and KOSEF (Korea);
CONICET and UBACyT (Argentina);
FOM (The Netherlands);
STFC and the Royal Society (United Kingdom);
MSMT and GACR (Czech Republic);
CRC Program, CFI, NSERC and WestGrid Project (Canada);
BMBF and DFG (Germany);
SFI (Ireland);
The Swedish Research Council (Sweden);
CAS and CNSF (China);
and the
Alexander von Humboldt Foundation (Germany).

\end{document}